\documentclass{article}
\usepackage{arxiv}
\usepackage{natbib}
\usepackage{graphicx}
\usepackage{multirow}
\usepackage{subcaption}
\usepackage{booktabs}
\usepackage{makecell}
\usepackage{todonotes}
\usepackage{lmodern,url}

\title{Combination of digital signal processing and assembled predictive models facilitates the rational design of proteins}
\usepackage{authblk}

\author[1,2\thanks{\tt{david.medina@cebib.cl}}]{David Medina-Ortiz}
\author[1,3]{Sebastian Contreras}
\author[1,4]{Juan Amado-Hinojosa}
\author[5]{Jorge Torres-Almonacid}
\author[1,4]{Juan A. Asenjo}
\author[6]{Marcelo Navarrete}
\author[1,4\thanks{\tt{aolivera@ing.uchile.cl}}]{\'Alvaro Olivera-Nappa}
\affil[1]{Centre for Biotechnology and Bioengineering, Universidad de Chile, Beauchef 851, Santiago, Chile.}
\affil[2]{Division of Chemistry and Chemical Engineering, California Institute of Technology, Pasadena, CA 91125, USA.}
\affil[3]{Max Planck Institute for Dynamics and Self-Organization, Am Faßberg 17, 37077 Göttingen, Germany.}
\affil[4]{Departamento de Ingenier\'ia Qu\'imica, Biotecnolog\'ia y Materiales, Facultad de Ciencias, F\'isica y Matem\'aticas, Universidad de Chile, Beauchef 851, Santiago, Chile.}
\affil[5]{Departamento de Ingenier\'ia En Computaci\'on, Universidad de Magallanes, Avenida Bulnes 01855, Punta Arenas, Chile.}
\affil[6]{Escuela de Medicina, Universidad de Magallanes, Avenida Bulnes 01855, Punta Arenas, Chile.}

\date{}

\renewcommand{\headeright}{ }
\renewcommand{\undertitle}{ }

\begin{document}
\maketitle

\begin{abstract}
Predicting the effect of mutations in proteins is one of the most critical challenges in protein engineering; by knowing the effect a substitution of one (or several) residues in the protein's sequence has on its overall properties, could design a variant with a desirable function. New strategies and methodologies to create predictive models are continually being developed. However, those that claim to be general often do not reach adequate performance, and those that aim to a particular task improve their predictive performance at the cost of the method's generality. Moreover, these approaches typically require a particular decision to encode the amino acidic sequence, without an explicit methodological agreement in such endeavor. To address these issues, in this work, we applied clustering, embedding, and dimensionality reduction techniques to the AAIndex database to select meaningful combinations of physicochemical properties for the encoding stage. We then used the chosen set of properties to obtain several encodings of the same sequence, to subsequently apply the Fast Fourier Transform (FFT) on them. We perform an exploratory stage of Machine-Learning models in the frequency space, using different algorithms and hyperparameters. Finally, we select the best performing predictive models in each set of properties and create an assembled model. We extensively tested the proposed methodology on different datasets and demonstrated that the generated assembled model achieved notably better performance metrics than those models based on a single encoding and, in most cases, better than those previously reported. The proposed method is available as a Python library for non-commercial use under the GNU General Public License (GPLv3) license.
\end{abstract}


\section*{INTRODUCTION}

Proteins are involved in nearly every known biological process, and by investigating protein structure and function it becomes possible to appreciate why they are integrated into living organisms \citep{goodey_understanding_2009}. The properties and functionalities of proteins are closely related to their constitutive amino acids, and consequently, to their physicochemical characteristics, which finally determine protein folding and biological structure \citep{sadowski_sequencestructure_2009}. Because of its intrinsically multidisciplinary nature, knowledge of the relationships between protein structure and function has vastly profited from advanced computational methods, structural biology,  biophysics, and other relevant disciplines \citep{poluri_rational_2017}.

Protein design finds industrial applications in the fields of chemical, biotechnological, and biomedical engineering. Examples thereof are the enzymes used to produce paper (xylanases, cellulases), starch processing (amylases, isomerases), detergents (proteases, lipases), textiles (peroxidases, cellulases, catalases), bioremediation (enzymes to degrade xenobiotic compounds), food and pharmaceuticals (pectinolytic enzymes, lipases, lactase), catalytic antibodies, among several other catalytic and non-catalytic applications \citep{poluri_world_2017}. For carrying out the challenging task of designing and generating the engineered proteins with tailored -enhanced- functions, researchers have mainly followed two approaches in the past two decades: directed evolution and rational designing \citep{poluri_world_2017}.

Directed evolution is a protein improvement technique set as an optimisation problem. It is based on identifying variants with desirable characteristics, selected through processes inspired by natural selection, becoming one of the essential techniques in the field of protein engineering in the last century. On the other hand, the rational design of proteins requires computational methods to identify protein sequences and predict their folding to specific conformations \citep{ding_emergence_2006}. Firstly, protein rational design approaches were mostly based on the proteins' primary sequence composition, excluding secondary or tertiary interactions \citep{poluri_rational_2017}. The improvement of algorithms and the creation of data repositories with vast information on amino acid's physicochemical properties has allowed rational protein design to predict protein structural conformation and functional characterisation more accurately. Such predictions are enhanced when combined with structural information, like X-ray crystallography and nuclear magnetic resonance spectroscopy (NMR) imaging. 

Different libraries and computational methods have been developed to support the rational design of proteins and optimize their directed evolution. Tools such as Site Directed Mutator (SDM) \citep{worth2011sdm}, FoldX \citep{schymkowitz2005foldx}, and AutoMute \citep{masso2010auto}, among others, have mainly focused on the evaluation of protein stability using thermodynamic analysis and structural bioinformatics approaches; however, they require highly specific structural information, and their computational cost can be high. Another types of tools are based on the use of phylogenetic properties to analyze the sequences, one of these being MOSST \citep{olivera2011mutagenesis}. Machine Learning (ML) strategies have been widely applied in protein engineering, whether for the design of mutations through directed evolution \citep{wu2019machine}, development of peptide classification systems, prediction of protein solubility \citep{10.1093/bioinformatics/bty166, 10.1093/bioinformatics/btx662, 10.1093/bioinformatics/btx345}, or evaluation of the effect of variations in clinical applications \citep{gossage2014integrated, usmani2018prediction, chen2016iacp}, among others. However, when designing a prediction system, there are several challenges, like adapting the information to train the predictive models, selecting algorithms, evaluating their performance, and  coding of amino acidic residues into numerical vectors.

Typical approaches are based on the One Hot encoding and Ordinal encoding strategies. However, these techniques do not consider the residues characteristics \citep{yang2018learned}. Recent advances in text mining have allowed applying doc2vec methods to develop embedding representations, being widely used in different studies \citep{sinai2017variational, xu2017idhss, greener2018design}. Nevertheless, the computational resources are higher in comparison with the previous strategies.

Physicochemical properties have been intensively used for the characterisation and encoding of amino acid sequences, and subsequent training of Machine-Learning (ML) models, being AAIndex \citep{kawashima2000aaindex} one of the available databases for such purposes. Properties listed in AAIndex are 566 to date, which result far too many for training ML models, and the selection of what properties to use for coding poses a pattern identification problem or dimensionality reduction \citep{georgiev2009interpretable}. Different approaches based on unsupervised learning algorithms have been implemented \citep{saha2012fuzzy, forghani2017multivariate}. Nevertheless, a relationship between the group members and their descriptions has not been precisely defined, generating groups with mixed properties \citep{georgiev2009interpretable}. 


Various studies have combined physicochemical properties and digital signal processing in protein engineering \citep{cosic1987prediction, HEJASEDETRAD2000149}. \citet{4122061} was the pioneer in proposing the use of the electron-ion interaction potential (EIIP) property, and the application of discrete Fourier transforms (DFT) to analyze DNA and protein sequences, stating that the characterisation of sequences employing Fourier transforms facilitate the identification of patterns in protein families representing particular or unique behavior associated with the function or property of the sequences. Considering the advantages exposed by \citet{4122061}, methods such as the Resonant Recognition Model have been developed to correlate effects of protein mutations with digital sequences and apply them in cancer studies \citep {Cosic2016}, analysis of conserved regions \citep{HEJASEDETRAD2000149}, and evaluation of bioactivity \citep{335859}, among others. This is an important support for proteo-chemometrics applications \citep{LAPINSH2001180}, demonstrating the usefulness of these strategies in different research fields. Recent studies propose using digital signal processing in Machine Learning applications \citep{cadet2018application, cadet2018machine}, managing to develop predictive models for the evaluation of enantioselectivity, stability analysis, among other variables of common interest. However, they do not have an efficient method of selecting physicochemical properties; and their predictive methods are based only on the application of algorithms based on Support Vector Machine (SVM) or Partial Least Square (PLS), not considering other types of algorithms or applications of assembled strategies to improve the performance of their predictors.


In order to improve the capability of understanding and studying proteins only from their sequence, we propose an assembled-model approach, which can be utilized for a wide variety of proteins, reaching a high performance in protein prediction and characterisation using only its sequence. For this, we identify meaningful physicochemical properties groups from the AAIndex database, employing Natural Language Processing (NLP) and unsupervised learning methods on properties descriptions, to characterize and use those properties groups as descriptors of amino acid sequences. Those meaningful groups of properties are later used to encode amino acid sequences. We also propose a methodology for designing and implementing predictive models with direct application in protein engineering, inspired by digital signal processing and assembled strategies, since this representation is advantageous when emulating the effect of one amino acid over others in a sequence. Our strategy was validated in various case studies and datasets, generally achieving better performance metrics than those currently reported and assuring a high minimal performance, demonstrating the generality, versatility, and technological potential of the proposed methodology with different  protein molecules, which has not been reached by other strategies. Finally, we enable the strategy in the Python library \texttt{FFT-Predict}, freely available for non-commercial purposes. We believe this work's new approach can represent a powerful tool for protein engineering, since these assembled-models expedite the  challenges of designing proteins with desirable properties for any biotechnology branch.

\section*{METHODS AND DEVELOPMENT STRATEGIES}

\subsection*{SELECTION OF PHYSICOCHEMICAL PROPERTIES FOR ENCODING}

Physicochemical properties were obtained from the AAIndex database \citep{kawashima2000aaindex}. First, the database was pre-processed to be converted into a proper format for handling, using scripts implemented in Python language programming v3.6. After processing, embedding codifications were obtained for the descriptions of properties using doc2vec techniques \citep{xie2016unsupervised} and the \texttt{Gensim} Python library \citep{rehurek_lrec}. These embedding were clustered using different unsupervised learning algorithms and hyperparameters combinations, explored using the \texttt{DMAKit-Lib} library \citep{dmakit}, selecting those exhibiting the highest Calinski-Harabasz indexes \citep{zhang2017supervision}. After generating the partition, each cluster of properties was analyzed, removing those which values did not normally distribute among the amino acids employing the Kolmogorov–Smirnov test. Afterward, it a semantic re-ordering of the clusters was performed, where properties of the same nature were put together to form a property group. Finally, Principal Component Analysis (PCA) techniques were used to select a representative element of each cluster's properties.

\subsection*{FFT-PREDICT LIBRARY IMPLEMENTATION}

The \texttt{FFT-Predict} library proposed in this work was designed under the Object-Oriented Programming paradigm \citep{wegner1990concepts}, which is advantageous for the encapsulation and modularity of the lybrary. Its implementation relies on a set of modules written in Python version 3.6. All modules for generating supervised learning models use the \texttt{Scikit-learn} \citep{scikit-learn} and \texttt{TensorFlow}  \citep{tavenard2020tslearn} libraries. Dataset management is performed using the \texttt{Pandas} library \citep{mckinney-proc-scipy-2010} and statistics of the process are getting employing \texttt{Numpy} library. Finally, the \texttt{FFT-Predict} library has been encapsulated in an Anaconda environment allowing easy installation and increasing its portability.

\subsection*{OVERVIEW}

The \texttt{FFT-Predict} library is composed of different modules, which allow the development of predictive assembled models for amino acid sequences based on digital signal processing and employing the groups of physicochemical properties proposed in this work. Here, we will describe the main components of the library and their particular operating characteristics in a general way.

\begin{figure*}[!htpb]
    \centering
    \includegraphics[width=\textwidth]{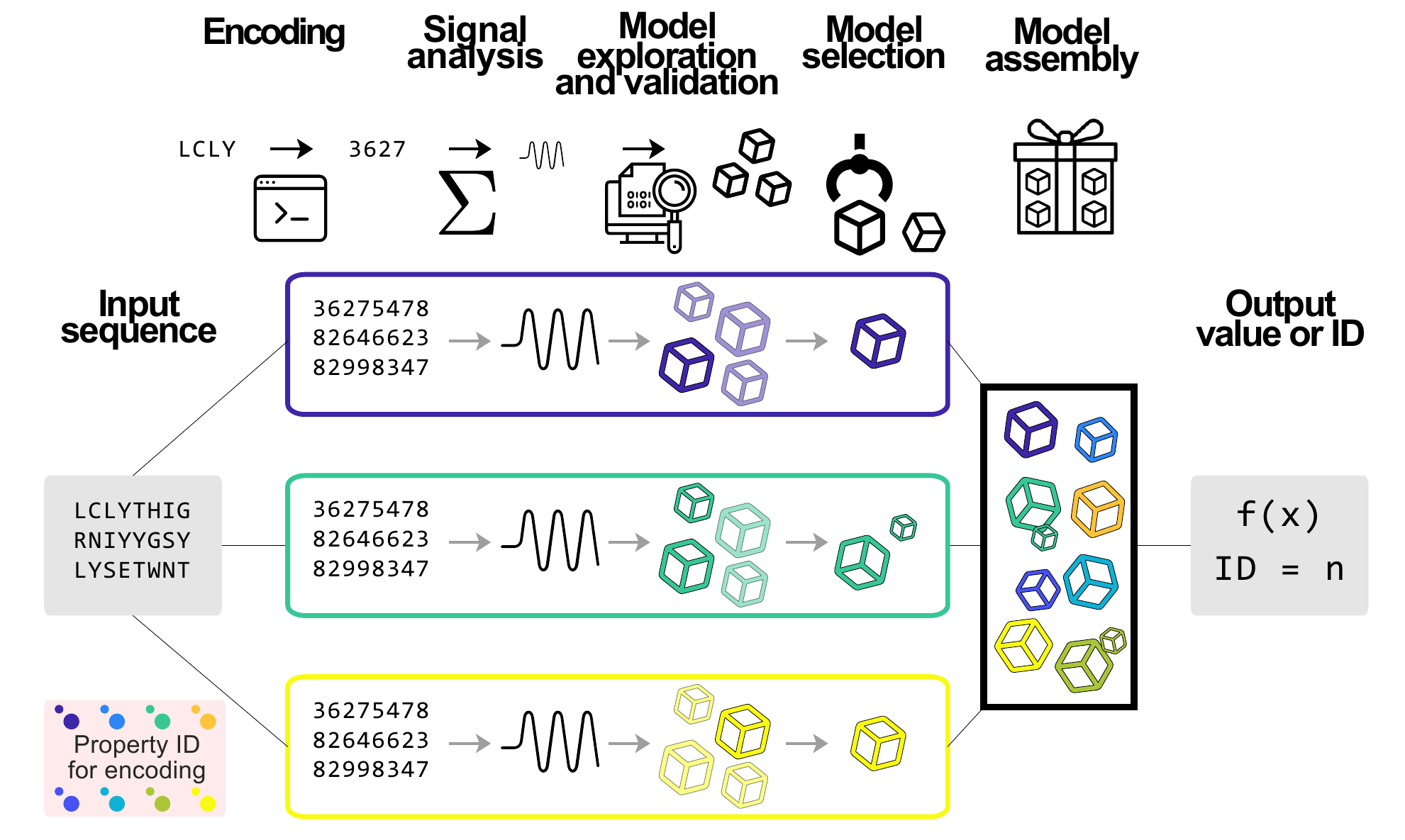}
    \caption{\textbf{Overview of the proposed methodology.}}
    \label{fig:schema}
\end{figure*}

\begin{itemize}

\item \textbf{Data pre-processing module}. In all input datasets, each example consists of amino acid sequences and a response variable, which may be related to catalytic activity, physiological or pathological parameters, thermodynamics, and stability, among other relevant properties. First, the library removes null values and encodes all sequences using the groups of physicochemical properties obtained applying the cluster methodology proposed in this work. After encoding, the digital signal processing is made over the encoding sequences using the Fast Fourier Transformation (FFT) algorithm \citep{cooley1970fast} for each group of property in combination with zero-padding techniques. Next, the Min-Max scaler is employed as a standardisation strategy for the datasets. Finally, the library splits the input dataset into two: The training dataset (80\%) and the testing dataset (20\%).

\item \textbf{Model exploration, training, validation, and selection module}. The training of the predictive models that conform the assembled-model includes those based on supervised learning algorithms implemented in the \texttt{DMAKit-Lib} library \citep{dmakit} and different Neural Networks architectures using the \texttt{TensorFlow} library \citep{abadi2016tensorflow}, exploring a total of 5382 different combinations of algorithms and hyperparameters in an exploration stage for each group of property (For more details see Table 3 of Supplementary Material). For all cases, we use $k$-fold cross-validation to prevent overfitting in the model training stage. Finally, different metrics are applied to evaluate the performance models, depending on the type of prediction to be made (See Table 4 of the Supplementary Material).

\item \textbf{Model-assembling module}. \texttt{FFT-Predict} Library selects the best combinations of algorithms and hyperparameters tested based on the performance measures obtained by applying statistical tests to identify outliers. The models for each group of physicochemical properties are stored in a pool of models and used to obtain predictions using the validation dataset. Finally, the library implements a weighting system to obtain the final predictive value compared with the real responses reported in the validation dataset, getting the assembled model performance metrics generated by the \texttt{FFT-Predict} Library.

\end{itemize}

\subsection*{DATASET SELECTION FOR TESTING AND CASE STUDIES}

Each dataset used in this article was selected to demonstrate the usability and applicability of \texttt{FFT-Predict} Library and the methodology proposed in this work. The datasets analyzed and tested in this work represent different protein engineering problems, such as evaluating the effect of variants in thermostability and enantioselectivity properties, classification of DNA-Binding Proteins, recognition of antimicrobial peptides, identification of quorum sense peptides, immunomodulatory peptide analysis, and protein solubility evaluation, among others. Each of these challenges have been used to inspire the development of predictive models or present experimental validations of their results. Thus, we obtained the datasets from databases reported in the literature. 

\subsubsection*{DNA BINDING PROTEIN}

Classification of DNA-binding proteins (DBP) is one of the most exciting biotechnology problems, mainly due to the direct applicability that its rational design has in protein engineering and synthetic biology to improve recombinant expression, and in molecular biology and genetic engineering, being directly applicable in the improvement of commercial DNA polymerases and restriction enzymes. Different computational methods have been proposed to develop DNA-Binding protein classification models, using various sequence coding and characterisation strategies and various predictive model training methodologies. Despite the enormous efforts, it is a problem that persists, and that demonstrates the difficulty of its resolution.

\subsubsection*{ENANTIOSELECTIVITY AND THERMOSTABILITY}

Thermostability is one of the most studied effects of mutations in protein engineering, since changes in the sequences that affect stability imply changes in their functionality, proving to be a sequence-dependent property \citep{worth2011sdm}. On the other hand, the analysis of stoichiometric effects, measured from enantioselectivity, is one of the most complex properties to predict, since it depends on the sequence and the preference for a type of enantiomer. Both properties represent a prediction problem. To verify the versatility and adaptability of the proposed methodology to predict continuous variables, we used the datasets of cytochrome P450 thermostability ('T50') \citep{li2007diverse} and of Epoxide hydrolase enantioselectivity \citep{yang2018learned}.

\subsubsection*{ANTIMICROBIAL PEPTIDE CLASSIFICATION}

Antimicrobial peptides (AMPs) are known as host defense peptides \citep{sitaram2002host}. These molecules play an essential role in the immune system's innate response, being of great interest due to their application in the pharmaceutical, biotechnological, and industrial areas \citep{papagianni2003ribosomally, ma2018improved}. Different computational methods based on Machine Learning strategies have been developed to classify antimicrobial peptides and their categorisation into various types \citep{xiao2013iamp, chen2016iacp, yi2019acp, zimmer2018artificial}. To demonstrate the adaptation of the method proposed in this work for datasets formed by non-protein sequences and to expose the advantages of digital signal processing, we use the datasets reported by \citet{xiao2013iamp} for the development of classification models of AMPs peptides. 

\section*{RESULTS AND DISCUSSION}

\subsection*{SELECTION OF PHYSICOCHEMICAL PROPERTIES FOR ENCODING}

\begin{figure*}[!htpb]
    \centering
    \includegraphics[width=12cm]{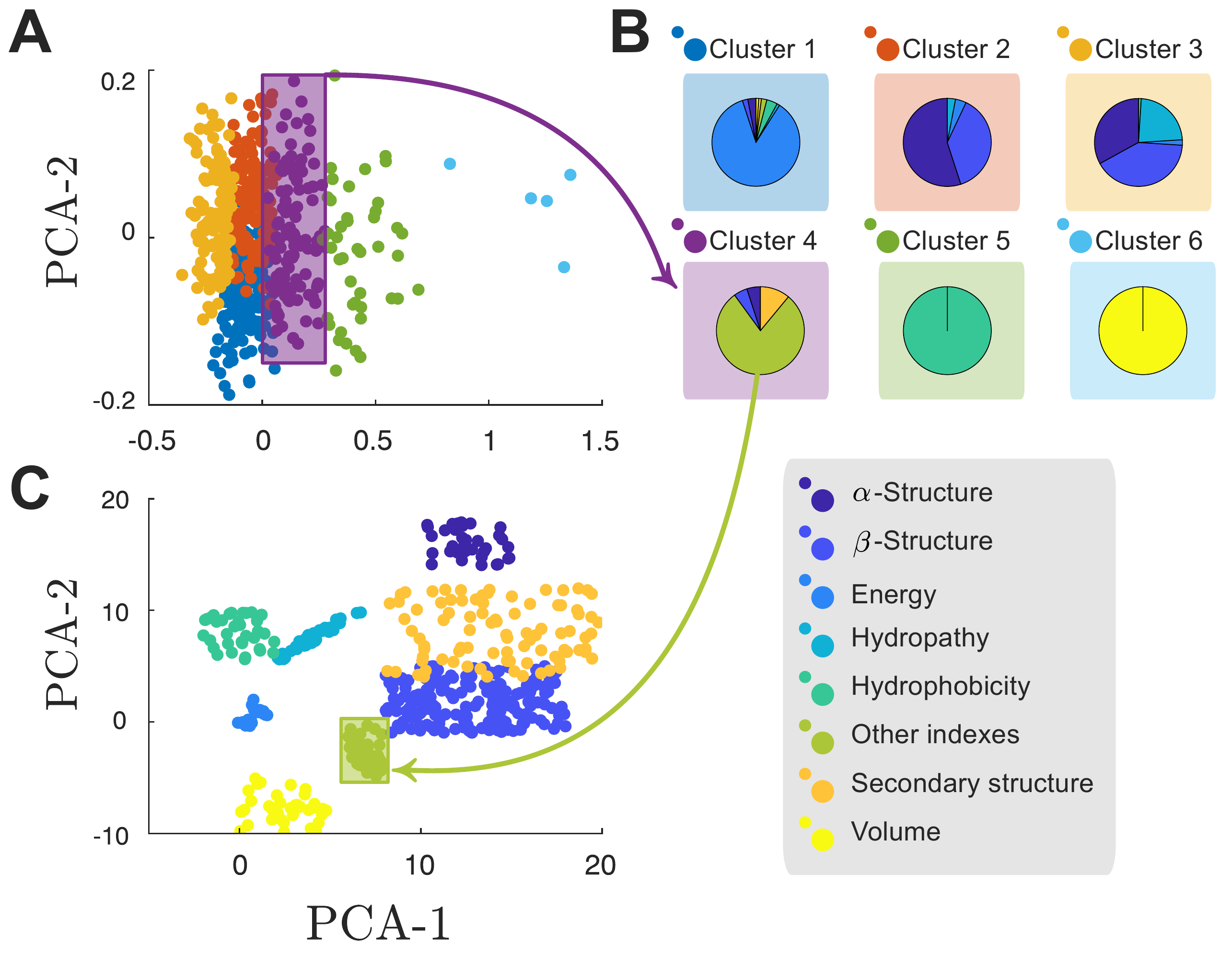}
    \caption{\textbf{Combination of classical unsupervised clustering with semantic search and Principal Component Analysis (PCA) allows the identification of linearly separate clusters of physicochemical properties of amino acids}. First, we explore different clustering methods and models to create groups of the physicochemical properties listed in AAIndex, to select the best based on its Calinski-Harabasz index.  In (\textbf{A}) we present the first two PCA components of the best clusterisation, which produced six clusters of properties (coloured accordingly) which, although well separated, presented an evident superposition between some cases. When semantically assessing the properties within the partition, eight different keywords appeared (\textbf{B}), inducing a subsequent re-assignment into 8 groups, which were linearly separated in the PCA-1/PCA-2 plane (\textbf{C}). We select the first component of each intra-cluster PCA for describing that keyword and for -independently- encoding the amino acid sequence to be analyzed.}

    \label{fig:fig1}
\end{figure*}

To identify groups of properties with a semantic sense, which are highly intuitive and generate an interpretation of the results from a mathematical point of view and a functional perspective of the properties, we propose a semantic clustering approach to combining NLP properties and unsupervised learning strategies. Five hundred and sixty-six descriptions of physicochemical properties extracted from the AAIndex database were encoded using embedding representations. The best partition in the exploration stage was obtained with the $k-$ means algorithm ($k=6$), achieving a Calinski-Harabasz index of 256.41 (Figure \ref{fig:fig1} A). The identified partitions were semantically evaluated to maintain a relationship between the members of a division and a property type, identifying eight keywords or property types, those related to secondary structure, energy and thermodynamics, hydrophobicity, volume, hydropathy, and different other indexes. The distribution of keywords in the six groups obtained by embedding was analyzed (Figure \ref{fig:fig1} B). The groups were reclassified based on the keywords, generating eight groups of physicochemical properties, which present broad semantic cohesion and allow an amino acid sequence to be described from different points of view (Figure \ref{fig:fig1}C). It is worth to mention that when forcing a partition based on a $k$-means algorithm with $k=8$ to match the final number of groups after the semantic re-ordering -- or any algorithm where the last number of groups is preset--, the resulting clusters were not semantically consistent, as none of the identified keywords would become evident when analyzing the generated partitions. Finally, each group of properties is represented by linear combinations in orthogonal vectors using PCA techniques, achieving at least 85\% of the total variance in the first component for each case (Table~\ref{tab:summary_properties}).  This way, the groups of properties identified can be used as coding or characterisation strategies for amino acid sequences, to subject them to algorithms to predict or identify patterns.

\begin{table}[!htpb]
\centering
\begin{tabular}{ccc c}
\toprule
Cluster-ID &
\textbf{Property Keyword}  &
\textbf{\# elements} &
\makecell[c]{\textbf{Variance}\\ ($1^{\rm st}$ Component)} \\ \midrule
ID-01 & $\alpha$ structure             & 37  & 95.95\% \\
ID-02 & $\beta$ structure               & 45  & 87.31\% \\
ID-03 & Energy  & 35  & 91.44\% \\
ID-04 & Hydropathy                                                                & 20  & 97.10\% \\
ID-05 & Hydrophobicity              & 27  & 91.73\% \\
ID-06 & Other indexes          & 14  & 89.85\% \\
ID-07 & Secondary structure           & 191 & 96.70\% \\
ID-08 & Volume                                                                    & 88  & 97.74\% \\ \bottomrule
\end{tabular}
\caption{Description of the groups of properties obtained applying the methodology proposed in this work to the AAIndex database.}
\label{tab:summary_properties}
\end{table}

\subsection*{PROPOSED WORKFLOW, TESTING, AND COMPARATIVE RESULTS}

In this work, we propose a methodology for designing and implementing predictive models with applications in protein engineering. The proposed method is inspired by the use of digital signal processing and assembly strategies. The datasets are made up of a group of sequences and a vector representing the variable to be predicted. Using the eight previously identified groups of physicochemical properties, the sequences are encoded. Fast Fourier Transform (FFT) is used to obtain digital signals from the encoded elements, thus preparing the model training datasets. In a second stage, different combinations of supervised learning algorithms and hyperparameters are explored and evaluated with varying measures of performance according to the type of response (see Table 3 and 4 of Supplementary Material). The best models are selected based on the application of statistical tests to identify outliers. In this way, for each group of physicochemical properties, predictive models are generated and stored in a pool of models that make up the assembled model. Finally, the predictions are generated based on a voting or weighting system analogous to bagging or boosting methods, and are compared with the real data, obtaining the performance measures of the generated assembled-model. The flow and the particular characteristics of the proposed method are shown in Figure~\ref{fig:schema}.

Different predictive assembled models were generated to evaluate the capacity of the method proposed in this work. Table \ref{tab:summary_task} describes the datasets, the prediction tasks, and the size of each input dataset used to assess the methodology proposed.

\begin{table*}[!htpb]
\centering
\begin{tabular}{l p{2cm} lcc}\toprule
\textbf{Dataset} & \textbf{Task} & \textbf{Property} &\textbf{Dimension} & \textbf{References} \\ \midrule
Enantioselectivity &\multirow{5}{*}{Prediction} &  Enantioselectivity    & 152   & \citep{yang2018learned}      \\
T50                &            &  Thermostability       & 261   & \citep{yang2018learned}      \\
Solubility         &            &  Protein Solubility    & 2985  & \citep{han2019develop}       \\
RT prediction       &           & \makecell[l]{Retention time \\  (Liq. Chromatography)} &  139073 &  \citep{ma2018improved} \\
PoP prediction     &            &  Peptide Observability & 95305 & \citep{zimmer2018artificial} \\\midrule
ACP-DL              &   \multirow{4}{*}{Classification} & Anticancer Peptides &  937 &  \citep{yi2019acp} \\
AntiTBP             &           & Antitubercular Peptides &  460 &  \citep{usmani2018prediction} \\ 
AMP-Binary          &           & Antimicrobial Peptides &  1832 &  \citep{sitaram2002host} \\
AMP-Multi           &           & Antimicrobial Peptides &  3796 &  \citep{sitaram2002host} \\
QSP                 &           & Quorum Sensing Peptides &  417 &  \citep{rajput2015prediction} \\ \bottomrule
\end{tabular}
\caption{Description of the datasets considered for the Case Study.}
\label{tab:summary_task}
\end{table*}

\begin{figure*}[!htpb]
    \centering
    \includegraphics[width=12cm]{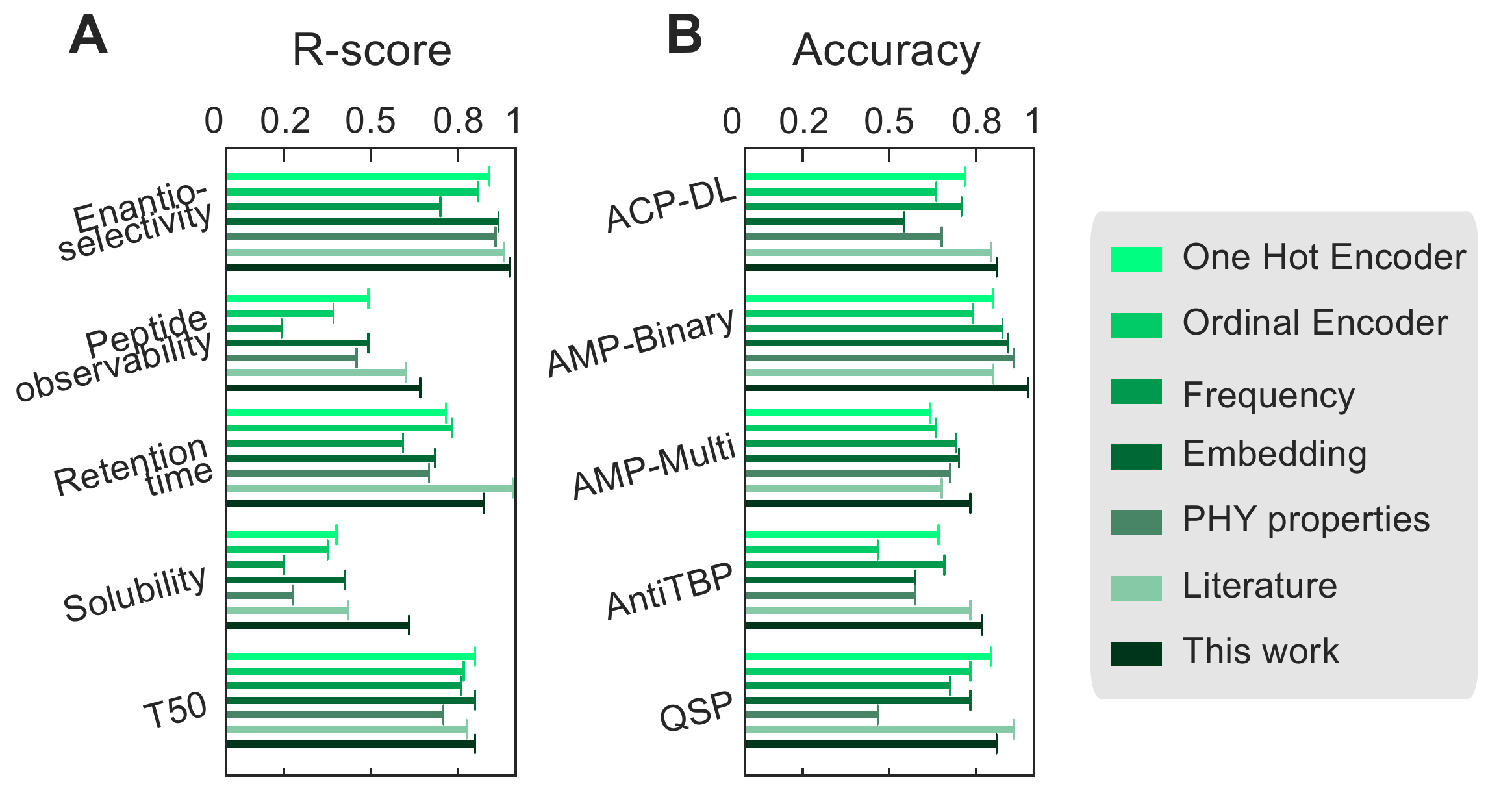}
    \caption{\textbf{Comparison of the performance metrics achieved by classical encoding methodologies, literature, and our work, for prediction (A) and classification (B) tasks.} We present best models' performances for each of the different encoding strategies evaluated (namely One Hot Encoder, Ordinal Encoder, Frequency, Embedding, physicochemical (PHY) properties), the highest performance reported in literature for the dataset, and our results. Description of the different datasets and tasks are presented in Table~\ref{tab:summary_task}. Our methodology achieves the highest performance metrics in most of the presented tasks (and if not, the second), highlighting its versatility and generality.}
    \label{fig:comparison_models}
\end{figure*}

Predictive assembled models were obtained for each dataset using the methodology proposed in this article. Different classical coding strategies such as One Hot Encoder, Ordinal Encoder, and residual frequency were used to compare the obtained performances inspired by digital signal processing. Besides, the TAPE tool \citep{rao2019evaluating} was used to obtain embedding representations of the sequences. Figure \ref{fig:comparison_models} summarizes the results obtained for the different encodings and the performances achieved by the method proposed in this article. Complete detail of the performances and their comparisons are shown in Tables 1 and 2 of the Supplementary Material. The results show that the proposed strategy achieves a better or equal performance than the current codings and performance previously reported with specific methods for each case. Only in particular circumstances, the performance obtained is slightly lower; however, the difference is not significant. This shows the generality of the method and its application to different cases and biotechnological problems, being reflected not only in the adaptation of the strategy to various case studies and applications, since it facilitates the prediction of properties that are not directly related to the sequence primary protein, such as DNA-Binding protein classification, estimation of enantioselectivity, prediction of quorum sense peptides, among other evaluated tasks, properties which depend on structural properties and the disposition of residues in space, as well as folding factors and thermodynamic effects that dictate the conformation of the protein.

It is possible to hypothesise that this generalisation and improvement of the assembled models' performance generated concerning different coding strategies or particular methods for specific cases, are achieved for two fundamental reasons. First, various physicochemical properties combined with digital signal processing allow efficient characterisation of amino acid sequences, since descriptors are considered at the sequence level, and the contribution of each residue to its neighbourhood is also considered, due to the characteristics of Fourier transforms \citep{Cosic2016, cosic1987prediction}. Second, the combination of supervised learning algorithms and configuration hyperparameters, statistically selected according to many performance measures for the generation of the assembled models, generates a synergy between them. Different points of view are considered simultaneously for developing predictive models. Besides, as different performance measures are considered, each model's particular characteristics are used to increase the predictive power of the generated assembled model \citep{medina2020development}.

\subsection*{CASE STUDY: DNA-BINDING PROTEINS (DBP)}

The binary classification assembled model generated using the methodology proposed in this work had 84.19\% accuracy and 84.96\% precision, achieving a significant improvement with respect to the models previously developed by \citet{rahman2018dpp}, \citet{WEI2017135}, and \citet{ADILINA201964} with 77.42\%, 79.00\%, and 82.26\% accuracy, respectively (See Table \ref{tab:compare_models} for more details). However, it was not possible to obtain the best performance compared to the previously reported methods since \citet{tan2020predpsd} achieved an accuracy of 91.20\%. Despite this, our method is practically the second-best, being a significant achievement for a generic strategy, applied to a specific and highly complex problem such as the DNA-Binding proteins classification, demonstrating the advantages of the combination of digital signal processing and assembled strategies for the development of predictive models.

\begin{figure}[htpb]
    \centering
    \includegraphics[width=10.5cm]{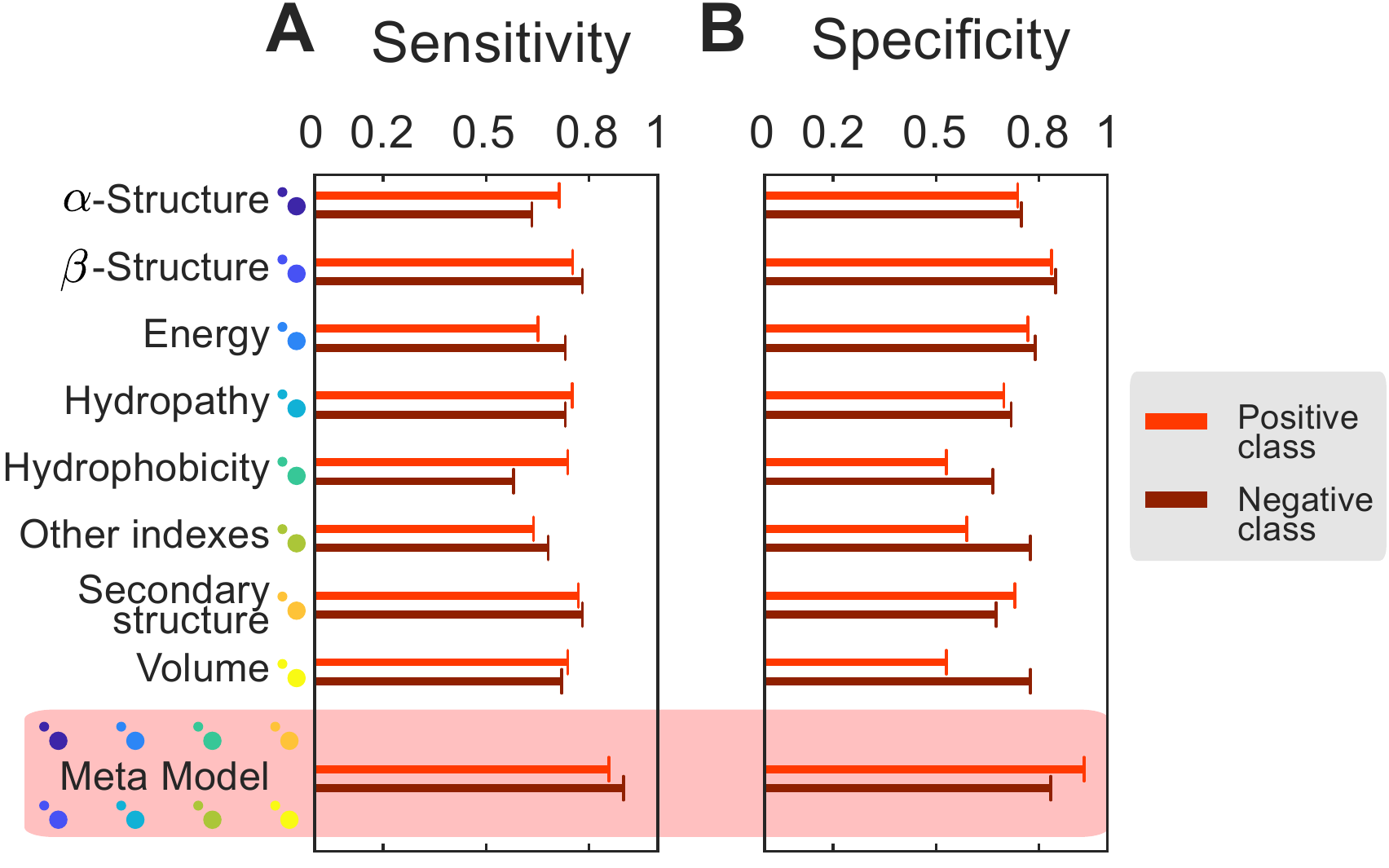}
    \caption{\textbf{The information provided by the different encodings is not redundant, as the assembled model's performance metrics are higher than those achieved by any of the other models.} Using the DNA-binding protein dataset, we encode the examples through the different proposed clusters of properties. Subsequently, we train predictive models for each one and evaluate their sensitivity (A) and specificity (B). We observe a significant increase in the performance metrics reached by the assembled model, higher than those reached by any of them, suggesting a synergistic effect that arises when combining different encodings.}
    \label{figsCase2}
\end{figure}

Sensitivity and specificity analyzes were performed on both the generated assembled model and its constituent models (Figure~\ref{figsCase2} A-B). Both metrics are lower for the individual models, compared to the assembled model. This shows that there is an improvement in the predictive power of the models promoted by the synergy that implies combining different properties to characterize the same set of sequences and the use of different algorithms and hyperparameters to solve the same problem.

\begin{table}[!htpb]
\centering
\begin{tabular}{l p{7cm} l}
\toprule
\textbf{Method} &  \textbf{Description} &  \textbf{Performance} \\ \midrule
DPP-PseAAC &  Using SVM as a supervised learning algorithm  and PseAAC as a method to characterized sequence &
  77.42\%  \citet{rahman2018dpp}\\
Local-DPP & Using RF as a supervised learning algorithm and  Pseudo Position-Specific Scoring Matrix as a method to characterized sequences &
   79.00\%  \citet{WEI2017135} \\
Chou’s general PseAAC &
  Grouped and Recursive Feature Selection based on Chou's  general PseAAC and Tree classifier as a prediction method &
  82.26\% \citet{ADILINA201964} \\
Network Similarity &
  Using network similarity as a method to create descriptors in dataset and CN as an ML algorithm &  83.50\% \citet{Wang2020} \\
PredPSD &  Use the minimal redundancy maximal relevance criterion (mRMR) feature selection algorithm into the gradient tree boosting (GTB) ML algorithm &  91.20\%  \citet{tan2020predpsd} \\
FFT-Predict &  Using physicochemical  properties digitized and a assembled strategy to developer predictive models based on a supervised learning algorithm &  84.19\%  This work \\ \bottomrule
\end{tabular}
\caption{Comparison of DNA-Binding protein predictive models previously reported with assembled model obtained using FFT-predict method}
\label{tab:compare_models}
\end{table}

\subsection*{CASE STUDY: ENANTIOSELECTIVITY AND THERMOSTABILITY PREDICTIONS}

\begin{figure}[htpb]
    \centering
    \includegraphics[width=10.5cm]{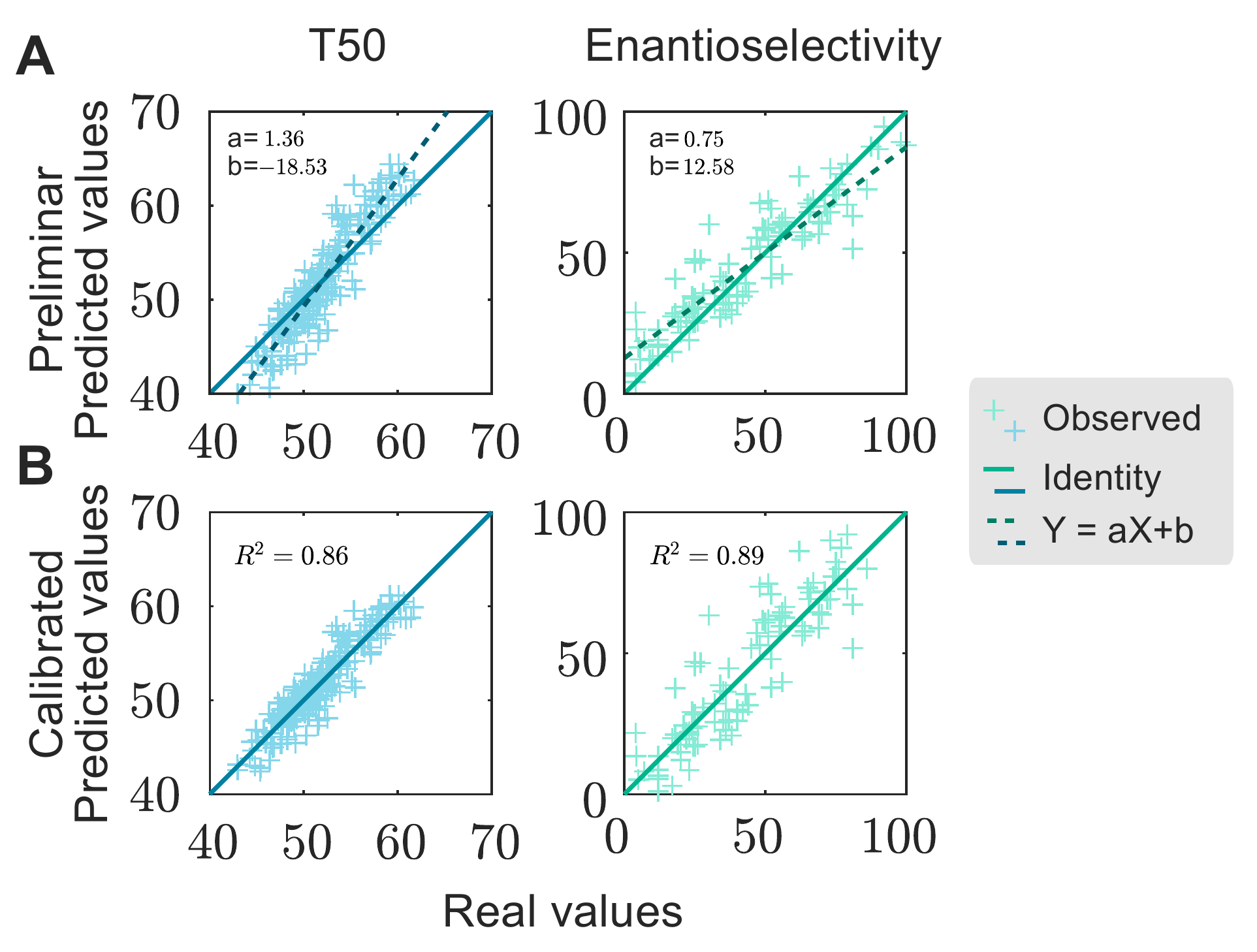}
    \caption{\textbf{Scatter plot of predictions versus real values for predictive models generated using the Enantioselectivity and T50 datasets, through the FFT-predict methodology.} Even though the produced models reached better performance metrics than those reported in literature, a re-calibration procedure might be necessary when the linear fit is off (from \textbf{A} to \textbf{B}).}
    \label{fig:case1}
\end{figure}

Assembled models for thermostability (T50) and enantioselectivity were successfully developed using the proposed methodology. The performance (Pearson's coefficient) OF both assembled models achieved corresponds -respectively- to 0.92 and 0.98, which is significantly higher than those previously reported obtained using different coding methods. In this way, we demonstrate that the approach proposed is highly suitable for protein variant datasets and facilitates the development of predictive models with high performance. Despite the high-performance measure obtained in both cases, an overfitting trend is observed in the assembled model, hence it is possible that a significant generalisation is not achieved due to the lack of data, especially for the enantioselectivity problem (See Figure~\ref{fig:case1} A). However, we propose a calibration method to adapt the predictions obtained and reduce the existing imbalance (See Figure~\ref{fig:case1} B). In this way, we update the performance measures, obtaining 0.94 for the case of enantioselectivity and 0.93 for thermostability, presenting a slight decrease for the first, while slightly increasing the second. Despite having generated a post-processing strategy of the assembled models, the performance measures were not significantly altered, which depicts the adaptability of the method to prediction of continuous variables, outperforming earlier digital signal processing strategies \citep{cadet2018application, cadet2018machine}.

\subsection*{CASE STUDY: ANTIMICROBIAL PEPTIDE CLASSIFICATION}

\begin{figure*}[htpb]
    \centering
    \includegraphics[width=15cm]{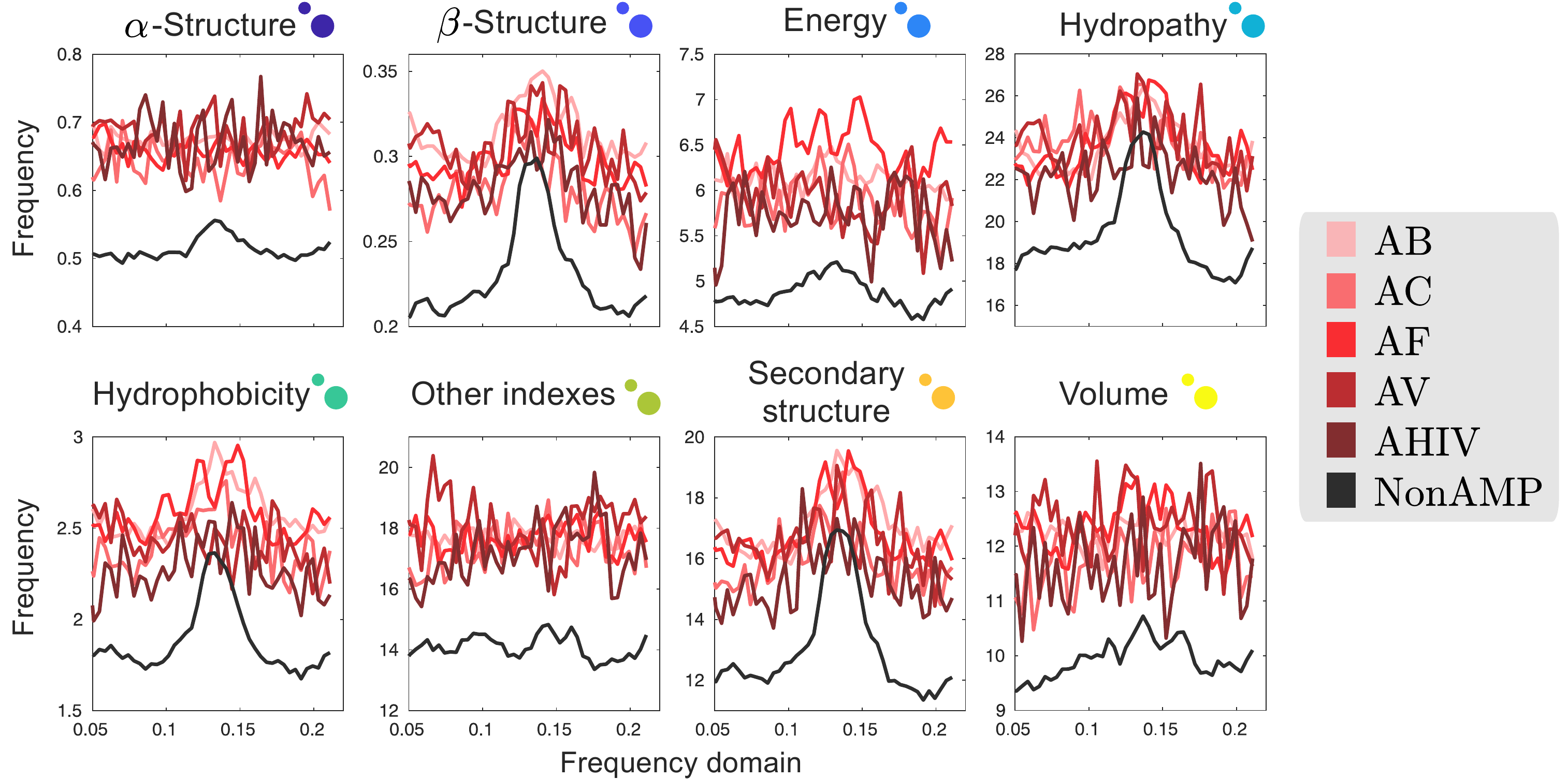}
    \caption{\textbf{Encoded amino acidic sequences are better separated in the frequency (Fourier) space, allowing us to produce models that reach higher performance}. Subfigures show the Fourier spectrum of different sequences of peptides, encoded according to the groups of properties proposed in this article. We analyze two types of peptides: Antimicrobial (AMPs) and non-Antimicrobial (nonAMP). AMPs are subsequently divided into five categories: Antibacterial Peptides (AB), Anticancer/tumor Peptides (AC), Antifungal Peptides (AF), Anti-HIV Peptides (AHIV), and Antiviral Peptides (AV). The signals analyzed show a clear differentiation for AMPs concerning nonAMPs.}
    \label{fig:fig1_case3}
\end{figure*}

Using the datasets reported by \citet{xiao2013iamp}, methods of identification of AMPs based on binary classification were successfully developed using coding by digital signal processing over groups of physicochemical properties proposed in this work and assembled strategies to demonstrate the capabilities and advantages of the method suggested in this article. AMPs peptides show a clear visual difference from non-antimicrobial peptides (nonAMP) (Figure~\ref{fig:fig1_case3}),  reflecting high accuracy (98.93 \%) and precision (97.28 \%) values obtained for the AMPs assembled model, denoting the advantages of this type of encoding and how it helps Machine Learning methods to identify patterns and learning of this type of classifiers.   

\section*{Conclusions}

In this work, we identify and extract meaningful groups of physicochemical properties from the AAIndex database to encode sequences of amino acids, and propose a novel methodology for creating predictive models using such encoded sequences. 

By combining Natural Language Processing with unsupervised learning algorithms, we explore and select the best partition of physicochemical properties within the AAIndex. The six resulting groups of properties, even though sufficiently separated, had an evident overlapping. By applying a semantic search over them, we identified eight keywords, which defined a subsequent partition that was linearly separated in the PCA1/PCA2 plane, accounting for the high semantic consistency among each group's properties within the partition. To summarize the information in each group, we performed an inter-group PCA, selecting as representative the first component of it, which in all cases explained more than the 85\% of the total variance.

Using the proposed 8-multi-property encoding, digital signal processing (Fourier transform), and assembly methods, we propose a general methodology for designing and implementing predictive models, with applications in protein engineering. First, the input dataset of amino acidic sequences is encoded, to be subsequently processed by Fast Fourier Transform. In the Fourier (frequency) space, followed by a model-exploration stage, where different combinations of machine-learning algorithms and hyperparameters are tried to select the best model(s), separately for each multi-property previously defined. Finally, all selected models are joined in a final assembly model, where the output delivered is produced by a weighted average/voting of the constitutive models. 

We test and evaluate the proposed method with different data sets for various prediction tasks with high relevance in the field of protein engineering, achieving always better performance metrics than classical approaches. Some exceptions were spotted in particular cases (when compared to highly specific methods), achieving in those very similar metrics to the best model. 

Remarkably, we found that when comparing different encodings into an assembled model, the predictions were far more accurate than those cast by individual models --in terms of performance metrics--, evidencing an exact synergic effect. Such synergy is not surprising; combining several models trained with different encodings allows them to solve the same task from different points of view, maximizing the assembled model's predictive performance and generalization.  

\section*{CODE AVAILABILITY}

All code is available at the authors' GitHub repository \url{https://github.com/dMedinaO/FFT_predict_Lib}.

\section*{SUPPLEMENTARY DATA}

Supplementary Data are available at the authors' GitHub repository \url{https://github.com/dMedinaO/FFT_predict_Lib}.

\section*{ACKNOWLEDGEMENTS}

This research has been financed mainly by the Centre for Biotechnology and Bioengineering - CeBiB (PIA project FB0001, ANID, Chile). DM-O gratefully acknowledges ANID, Chile, for Ph.D. fellowship 21181435. JA-H gratefully acknowledges ANID, for Ph.D. fellowship 21182109. SC received support from the Max-Planck-Society. MN and JT-A gratefully acknowledge ANID, Chile for Fondecyt 1180882 project and Universidad de Magallanes for MAG1895 project.

\subsubsection*{Conflict of interest statement.} None declared.


\end{document}